\documentclass[]{aa}
\usepackage[dvips]{graphicx}
\usepackage{natbib}
\usepackage{txfonts}
\def\ecs{\mbox{~erg~cm$^{-2}$~s$^{-1}$}}

\def\ecsa{\mbox{~erg~cm$^{-2}$~s$^{-1}$~{\AA}$^{-1}$}}
\def\ebv{\mbox{$E_{B-V}$}}
\def\lya{\mbox{Ly$\alpha$}}

\def\halpha{\mbox{H$\alpha$}}
\def\hbeta{\mbox{H$\beta$}}
\def\hgamma{\mbox{H$\gamma$}}
\def\peryr{\mbox{$\>\rm yr^{-1}$}}

\def\subsun{\mbox{$_{\normalsize\odot}$}}
\def\lesssim{\mathrel{\hbox{\rlap{\hbox{\lower4pt\hbox{$\sim$}}}\hbox{$<$}}}}
\def\gtrsim{\mathrel{\hbox{\rlap{\hbox{\lower4pt\hbox{$\sim$}}}\hbox{$>$}}}}

\begin{document}

\title{Abundances and kinematics of a candidate sub-damped Lyman$\alpha$
  galaxy toward PHL~1226\thanks{Based on observations obtained at the
    German-Spanish Astronomical Center, Calar Alto, operated by the
    Max-Planck-Institut f\"ur Astronomie Heidelberg jointly with the Spanish
    National Commission for Astronomy}}

\titlerunning{The candidate sub-DLA galaxy toward PHL~1226}

\author{L. Christensen\inst{1}
  \and R.~E. Schulte-Ladbeck\inst{2}
  \and S.~F.~S\'anchez\inst{1}
  \and T.~Becker\inst{1}
  \and K.~Jahnke\inst{1}
  \and A.~Kelz\inst{1}
  \and M.~M. Roth\inst{1}
  \and L.~Wisotzki\inst{1,3}
   }
\institute{Astrophysikalisches Institut Potsdam, An der Sternwarte 16, 
14482 Potsdam, Germany
 \and Department of Physics \& Astronomy, University of Pittsburgh, 3941 O'Hara Street, Pittsburgh, PA 15260,  USA
\and Institut f\"ur Physik, Universit\"at Potsdam, Am Neuen Palais 10, 14469 Potsdam, Germany
}

\mail{lchristensen@aip.de}
\date{Received 5 July 2004 / Accepted 8 September 2004} 
       
\abstract{ The spectrum of the quasar \object{PHL~1226} is known to have a
  strong \ion{Mg}{ii} and sub-damped Lyman$\alpha$ (sub-DLA) absorption line
  system with $N(\ion{H}{i})=(5\pm2)\times10^{19}$~cm$^{-2}$ at $z=0.1602$.
  Using integral field spectra from the Potsdam Multi Aperture
  Spectrophotometer (PMAS) we investigate a galaxy at an impact parameter of
  6\farcs4 which is most probably responsible for the absorption lines.  A
  fainter galaxy at a similar redshift and a slightly larger distance from the
  QSO is known to exist, but we assume that the absorption is caused by the
  more nearby galaxy.  From optical Balmer lines we estimate an intrinsic
  reddening consistent with 0, and a moderate star formation rate of
  0.5~M\subsun\peryr\, is inferred from the \halpha\ luminosity.  Using
  nebular emission line ratios we find a solar oxygen abundance
  $12+\log\textrm{(O/H)}=8.7\pm0.1$ and a solar nitrogen to oxygen abundance
  ratio $\log\textrm{(N/O)}=-1.0\pm0.2$.  This abundance is larger than those
  of all known sub-DLA systems derived from analyses of metal absorption lines
  in quasar spectra. On the other hand, the properties are compatible with the
  most metal rich galaxies responsible for strong \ion{Mg}{ii} absorption
  systems.  These two categories can be reconciled if we assume an abundance
  gradient similar to local galaxies.  Under that assumption we predict
  abundances $12+\log\textrm{(O/H)}=7.1$ and $\log\textrm{(N/O)}=-1.9$ for the
  sub-DLA cloud, which is similar to high redshift DLA and sub-DLA systems.
  We find evidence for a rotational velocity of $\sim$200~km~s$^{-1}$ over a
  length of $\sim$7 kpc.  From the geometry and kinematics of the galaxy we
  estimate that the absorbing cloud does not belong to a rotating disk, but
  could originate in a rotating halo.
   }
\maketitle

\section{Introduction}
Metal absorption line systems in QSO spectra are frequently used to derive
abundances of intervening neutral gas clouds of high column density.  A
classic definition divides strong QSO hydrogen absorption systems into
different regimes: Lyman limit systems (LLS) with column densities of neutral
hydrogen in the range $10^{17}<N(\ion{H}{i})<2\times10^{20}$~cm$^{-2}$, and
damped Lyman$\alpha$ systems (DLAs) having
$N$(\ion{H}{i})$~>~2~\times10^{20}$~cm$^{-2}$ \citep{wolfe86}. Sub-DLA systems
with $10^{19}~<~N(\ion{H}{i})~<~2\times10^{20}$~cm$^{-2}$ also show broad
absorption line wings characteristic for DLAs and contain a large fraction of
the neutral hydrogen at high redshifts \citep{peroux03}. Strong hydrogen
absorption lines in QSO spectra are believed to arise in gas-rich
(proto)-galaxies, but any relation between absorption line systems and the
galaxies responsible for these is not well understood. Deep imaging studies of
the fields around QSOs which exhibit strong \lya\ absorption lines have
revealed several absorbing galaxy candidates close to the line of sight
\citep[e.g.][]{lebrun97}.  Follow-up spectroscopy is necessary to confirm the
candidates as the corresponding absorbing galaxies.

To date only few (13) low-redshift DLAs and sub-DLA galaxies have been
identified \citep[][and references therein]{turnshek01,rao03,chen03,lacy03},
while at high redshift even fewer confirmations exist
\citep{moller93,djorgovski96,fynbo99,moller02,christensen04,moller04}.
Considering the difficulties of confirming the absorbing galaxies, alternative
approaches have been carried out to determine which types of galaxies produce
DLA systems.  Through measurements of relative abundances of various elements,
it has been suggested that DLA galaxies are likely dwarf galaxies
\citep{pettini99,prochaska03,dessauges03,nissen04}.  Sub-DLAs show the same
spread in metallicity as the DLA systems indicating a similar nature of the
two samples \citep{peroux03}.  Although metallicities of DLA and sub-DLA
systems can be easily derived from the QSO spectra, the metallicity of the
parent galaxy has only been determined for one DLA galaxy previously
\citep{schulte04}.

Studies have shown that \ion{Mg}{ii} absorption systems arise in halos of a
wide range of galaxy types \citep{bergeron91,steidel94} and are typically
associated with LLS. \ion{Mg}{ii} systems have been suggested to be related to
rotating disks \citep{charlton98}, while \citet{steidel02} found that
\ion{Mg}{ii} systems cannot be explained by simple disk models and suggested
that the absorption occurs in rotating halos.

The quasar \object{PHL 1226} (\object{Q0151+045} at $z=0.404$) has a strong
\ion{Mg}{ii} absorption system at $z=0.1602$ \citep[][hereafter
B88]{bergeron88}.  A column density of
$N(\ion{H}{i})~=~(5~\pm~2)\times10^{19}$~cm$^{-2}$ was measured in a UV HST
spectrum classifying the object as sub-DLA (Rao 2004, private communication).
A bright $V~=~19.2\pm$0.3 galaxy at a projected angular separation of 6\farcs4
to the west of the QSO was identified as being responsible for the absorption
system by B88, who named this galaxy G4.  For a flat cosmological model with
$\Omega_{\Lambda}=0.7$ and $H_0=70$~km~s$^{-1}$~Mpc$^{-1}$, which we use
throughout the paper, an impact parameter of 6\farcs4 corresponds to 17.6 kpc.
An additional fainter galaxy, G3 ($V=20.5$) at a similar redshift, was found
at an impact parameter 10\farcs9 to the north of the QSO.  Either galaxy could
be responsible for the absorption system considering the scaling-law between
\ion{Mg}{ii} halo sizes and galaxy luminosities
\citep{bergeron91,lebrun93,guillemin97}. Optical spectroscopy of G3 and G4 is
presented in \citet{ellison04}.

In this paper we present a study of the galaxy G4 using integral field
spectroscopy with the Potsdam Multi Aperture Spectrophotometer
\citep[PMAS][]{pmas00}.  We describe the data reduction in
Sect.~\ref{data_red}.  Spectra and synthetic narrow-band emission line images
are presented in Sect.~\ref{spec} From the spectra we derive the intrinsic
reddening, oxygen and nitrogen abundances, and star formation rate in
Sect.~\ref{prop_deriv}.  We analyse the kinematics of the galaxy itself in
Sect.~\ref{kin}. A discussion is presented in Sect.~\ref{disc}, and the
conclusions in Sect.~\ref{conc}.

\section{Observations and data reduction}
\label{data_red}
PMAS is an Integral Field Instrument developed at the Astrophysikalisches
Institut Potsdam. It is currently installed at the Calar Alto Observatory 3.5m
Telescope in Spain and has been available as a common user instrument since
July 1, 2002. Observations using integral field units have the advantage of
combining imaging and spectroscopy information in one dataset.  In 2002, we
started a project to use PMAS for the study of sub-DLA and DLA galaxies. The
great advantage of using integral field spectroscopy for the investigation of
the very few currently known sub-DLA and DLA galaxies is that they allow us to
obtain data on the galaxies' positions, velocity fields and star-forming
properties all in one data cube.  Such data are critical to derive the nature
of the galaxies causing absorption systems in QSOs. It is our goal to use this
new instrument to identify galaxies responsible for low- as well as
high-redshift QSO absorbers, and to study their properties.
  
During the pilot observing run for this project, we observed two QSO fields.
Observations for the DLA galaxy toward Q223+131 revealed an extended
Lyman-$\alpha$ emission nebula surrounding the galaxy responsible for the DLA;
the analysis was presented in \citet{christensen04}. A second object observed
in the pilot study was PHL~1226 (Q0151+045).  The 8\arcsec$\times$8\arcsec\ 
PMAS field was targeted at the galaxy G4 because it is closer to the QSO than
G3, and thus the more likely absorbing galaxy.  An additional criterion was
technical -- G4 is more than a magnitude brighter than G3 which is farther
than 8\arcsec\ away from PHL~1226; its observation would have required another
set-up and a large amount of observing time. Here, we report in detail on the
results which can be obtained from the PMAS data of G4.

We obtained 4$\times$1800s exposures of G4 divided over two nights on Sep. 5
and 8, 2002, using a grating with 300 lines~mm$^{-1}$ which resulted in a
spectral resolution of 6.6~{\AA} FWHM.  The chosen grating angle allowed to
cover the wavelength range 4575--7880~{\AA}. The first night was photometric
with a seeing of 1\arcsec, while the seeing was varying during the second
night and the conditions were not photometric.  A log of the observations is
presented in Table~\ref{tab:log}.

The PMAS instrument has two cameras; one for the spectrograph, and an
additional camera used for acquisition and guiding (A\&G camera) which is
equipped with a 1k$\times1$k SITe CCD.  Using data from the A\&G camera, one
can estimate the variations and the evolution of the sky quality during the
night. Photometry of the guide star images taken during the spectral
integrations, show variations with a standard deviation presented in column 5
of Table~\ref{tab:log}.

The spectrograph is coupled by 256 fibers to a 16$\times$16 element micro-lens
array. During the observations each lens covered 0\farcs5$\times$0\farcs5 on
the sky.  The detector is a 2k$\times$4k SITe ST002A CCD which was read out in
a 2$\times$2 binned mode.

\begin{table*}
\begin{tabular}{lllllll}
\hline \hline
\smallskip
date (UT)  & exposure time (s)& airmass  &seeing&
$\sigma_{\mathrm{star}}$(\%)  & S/N \\
\hline
 Sep. 5 2002 02:48 &   1800 & 1.19 & 1\arcsec          & 0.5  & 5.9\\
 Sep. 5 2002 03:20 &   1800 & 1.19 & 1\arcsec          & 0.5  & 5.7\\
 Sep. 8 2002 00:03 &   1800 & 1.57 & 1\farcs3--1\farcs5 & 10   & 3.9\\
 Sep. 8 2002 00:36 &   1800 & 1.14 & 1\farcs3--1\farcs5 & 9    & 4.2\\
\hline
\end{tabular}
\caption[]{Log of the observations. $\sigma_{\mathrm{star}}$ denotes
  the standard deviation of flux in images of the guide star during each
  integration. Column 6
  lists the signal to noise ratio of a one-dimensional spectrum within
  a 1\arcsec\ radius of the QSO.}
\label{tab:log}
\end{table*}

Data reduction was performed with IDL based routines written specifically for
PMAS data \citep{becker02}. After bias subtraction the spectra were extracted
using information of the location of the 256 spectra on the CCD obtained from
a calibration frame obtained with an exposure of a continuum emission lamp
immediately before or after each target exposure. In the extraction a Gaussian
line profile was used to increase the signal-to-noise ratio of the extracted
spectra. Wavelength calibration was done using calibration spectra of Hg-Ne
emission line lamps also obtained following the science exposures. The
accuracy of the wavelength calibration was checked using sky emission lines,
showing a standard deviation of 0.3~{\AA}. Corrections for varying
fiber-to-fiber transmissions as a function of wavelength uses flat field
spectra obtained from the twilight sky.

\begin{figure}
\centering
\resizebox{\hsize}{!}{\includegraphics[]{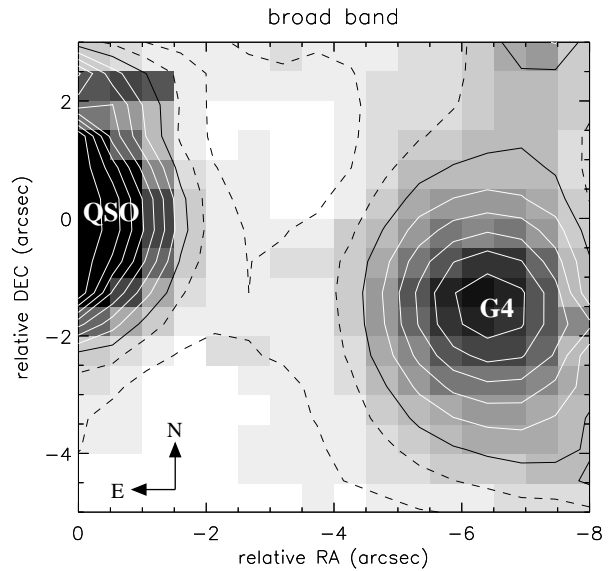}}
\caption{Broad band image in the wavelength range
  $4500~<~\lambda~<~8000$~{\AA} of the QSO at the position (0,0) and the galaxy
  G4 at $(-1.5,-6.5)$ as indicated.  Each spaxel represents a spectrum and the
  field of view is 8\arcsec$\times$8\arcsec. North is up and east is left.
  Coordinates are relative to the QSO position. }
\label{fig:br_map}
\end{figure}

Cosmic rays were removed from the spectra using the L.A. Cosmic routine within
IRAF \citep{vandok01}.  Each of the 4 exposures were corrected for an average
extinction value appropriate for Calar Alto before combining them.  The effect
of differential atmospheric extinction was corrected using the theoretical
approach described in \citet{filippenko82}. We checked whether the data cube
was appropriately corrected for the differential atmospheric refraction by
cross-correlating each monochromatic image with the broad-band image shown in
Fig.~\ref{fig:br_map}. This showed that the relative shift with wavelength was
smaller than 0\farcs08, which was negligible for the further analysis.
Subtraction of the sky background was done by creating an average sky spectrum
by selecting spaxels (spatial elements) located between the QSO and the
galaxy, uncontaminated by flux from any of these two.  This sky spectrum was
subtracted from all 256 spectra.

The method for combining the individual exposures was as follows. Firstly,
monochromatic images were made from the data cube at some selected wavelength.
Then we found a scale factor between the images, by calculating the total flux
in the images.  Because the spectra taken on Sep. 8 were of poorer quality
than the first ones, the individual data cubes were scaled to the ones from
Sep 5., and the final combination took the varying seeing into account by
applying a weighting scheme, where the weights were given from the signal to
noise ratio of each spectrum.  The signal-to-noise ratio was found from a 1D
spectrum created by co-adding spectra within 1\arcsec\ radial aperture
centered on the galaxy. These S/N ratios are listed in column 6 in
Table~\ref{tab:log}.  Finally, flux calibration was done the standard way by
comparing the spectra obtained of the spectrophotometric standard,
\object{BD+28$^\circ$4211} observed on Sep. 5 with table values.  After the
data reduction the spectra are contained in a data cube of dimensions
$16\times16\times1008$ pixels.

Further analysis of the reduced data cube was done using the Euro3D
Visualization Tool \citep{sanchez04a}, while one-dimensional spectra were
analysed using both IRAF and our own software (S\'anchez et al., 2004, in
prep.).  Fig.~\ref{fig:br_map} presents a composed broad-band image, with
dimensions 8\arcsec$\times$8\arcsec.  To the east (left), the QSO is seen at
the edge of the field. The absorbing galaxy G4, identified by B88, is the
principal source in the field at the position $(-1.5,-6.5)$ in agreement with
the impact parameter of 6\farcs4.

\section{Galaxy and QSO spectra}
\label{spec}
\begin{figure}
\centering
\resizebox{\hsize}{!}{\includegraphics[bb=50 0 820 453, clip]{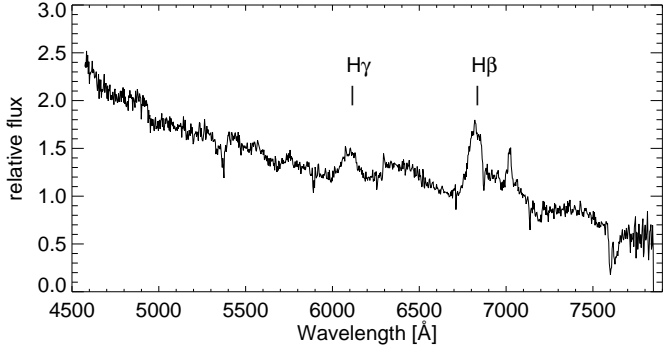}}
\caption{Spectrum of PHL 1226 created by co-adding 20 spaxels. The flux levels are relative since the
  QSO is located at the edge of the field of view. Broad emission lines from
  \hgamma\ and \hbeta\ at $z=0.404$ can be seen at 6100 {\AA} and 6820 {\AA},
  respectively.}
\label{fig:qso_spec}
\end{figure}

In the dust maps of \citet{schlegel98} a Galactic reddening of \ebv~=~0.051 is
found in the direction toward PHL 1226. The reduced spectra were de-reddened
using this value before deriving other quantities. From the reduced data cube,
one can create spectra from chosen spaxels or monochromatic- and narrow band
images in any wavelength region. A spectrum of the QSO created by co-adding 20
spaxels is shown in Fig.~\ref{fig:qso_spec}.

One sees that strong telluric absorption lines are present around 7600~{\AA}
which coincidentally corresponds to \halpha\ at the redshift of the absorber.
When line fluxes are derived, this feature will skew the results towards lower
values if not corrected for. Thus, to correct for this effect, a model for the
telluric absorption line was created. The QSO spectrum was smoothed using a
Gaussian function with $\sigma=10$~{\AA}, normalised to 1 at 7580 {\AA} and
outside the region around 7600 {\AA} the value was set to 1.  Following, all
spectra were divided by this model to correct for the absorption as
demonstrated in in Fig.~\ref{fig:tellur}, which brought out the presence of
the [\ion{N}{ii}] (6548~{\AA}) line in the spectrum of G4 as explained in
detail in Sect.~\ref{high_surf}.

\begin{figure}[t!]
\centering
\resizebox{\hsize}{!}{\includegraphics[bb= 54 -80 490 688,clip,angle=90]{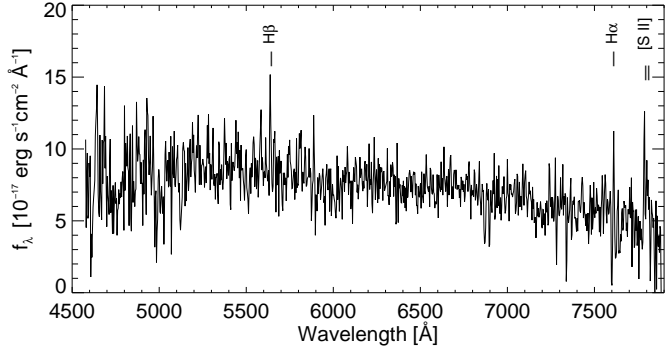}}
\caption{Spectrum of the galaxy created by co-addition of 112 spaxels. Any
  emission line present in the galaxy has been smeared out because many
  spaxels do not contribute to the line emission as shown in
  Sect.~\ref{narrow_band}. The seemingly strong [\ion{S}{ii}] lines are partly
  caused by errors in sky subtraction.}
\label{fig:gal_all}
\end{figure}

\begin{figure}[t!]
\centering
\resizebox{\hsize}{!}{\includegraphics[bb= 0 -546 290 -40,clip,angle=90]{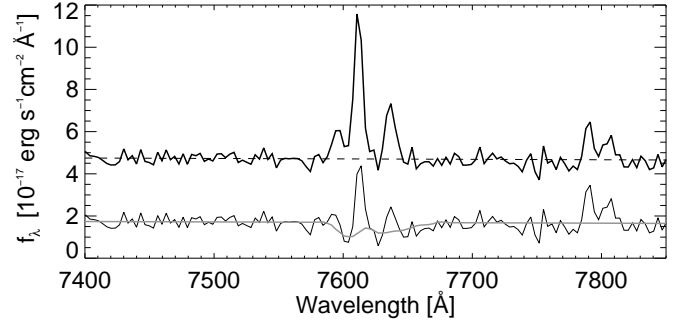}}
\caption{Zoom in of the spectrum in Fig.~\ref{fig:gal_spec} in the region
  around [\ion{N}{ii}], \halpha, and [\ion{S}{ii}]. This plot shows the effect
  of correcting for the telluric absorption feature around 7600~{\AA}. The
  lower spectrum (thin line) is the uncorrected spectrum, and the gray line is
  the model of the telluric feature created from the QSO spectrum. The upper
  spectrum (thick line) has been corrected for the absorption feature and
  offset by +3$\times10^{-17}$ \ecsa\ for clarity.}
\label{fig:tellur}
\end{figure}

Co-adding all spaxels (112 in an area of 4\arcsec$\times$7\arcsec)
corresponding roughly to the size of the galaxy in this dataset results in the
spectrum presented in Fig.~\ref{fig:gal_all}. This spectrum gives the total
continuum flux from the galaxy with S/N~=~6 and does not show strong
emission lines, specifically not before the correction for telluric absorption
is applied. One must note that co-adding 112 spaxels dilutes any emission line
signal coming from a potentially smaller region. Additionally, residuals from
the background subtraction can artificially enhance emission lines. This
appears to be the case for the sulfur lines in Fig.~\ref{fig:gal_all}, because
they lie close to strong sky lines.  In order to increase the signal to noise
ratio of the emission line spectrum, and be less affected by sky subtraction
residuals, a localization of the emission line region needs to be done.

To check the spectrophotometry, we convolved the galaxy spectrum with a
transmission curve of the Bessell $V$ band filter. We find $V=19.08\pm$0.05,
corresponding to an absolute magnitude $M_V$ = --20.3 for the galaxy in
agreement with the photometry in B88.

\subsection{Emission line images}
\label{narrow_band}
\begin{figure*}[t!]
\begin{minipage}[c]{.33\textwidth}
\resizebox{\hsize}{!}
{\includegraphics[]{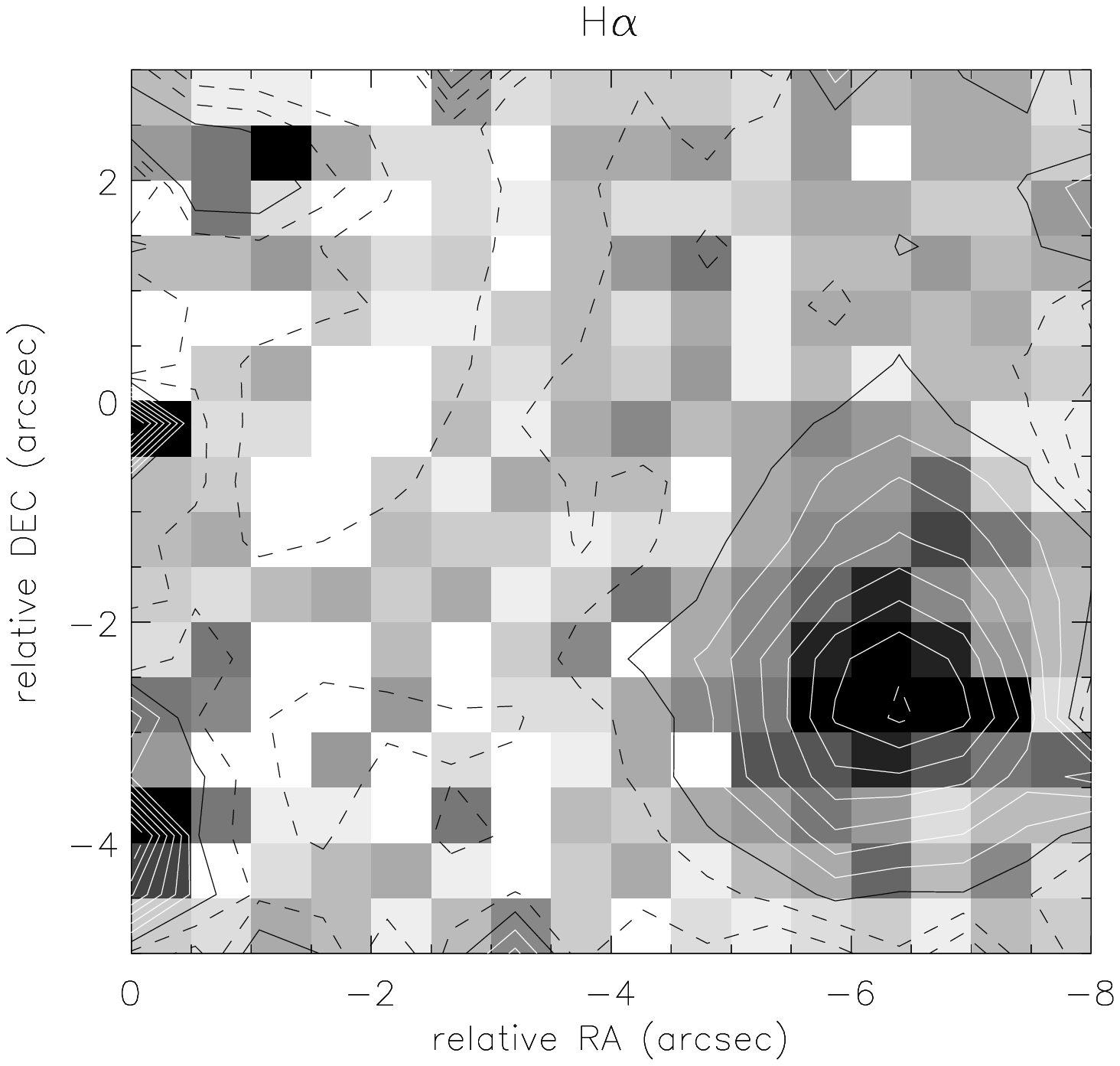}}
\end{minipage}%
\begin{minipage}[c]{.33\textwidth}
\resizebox{\hsize}{!}
{\includegraphics[]{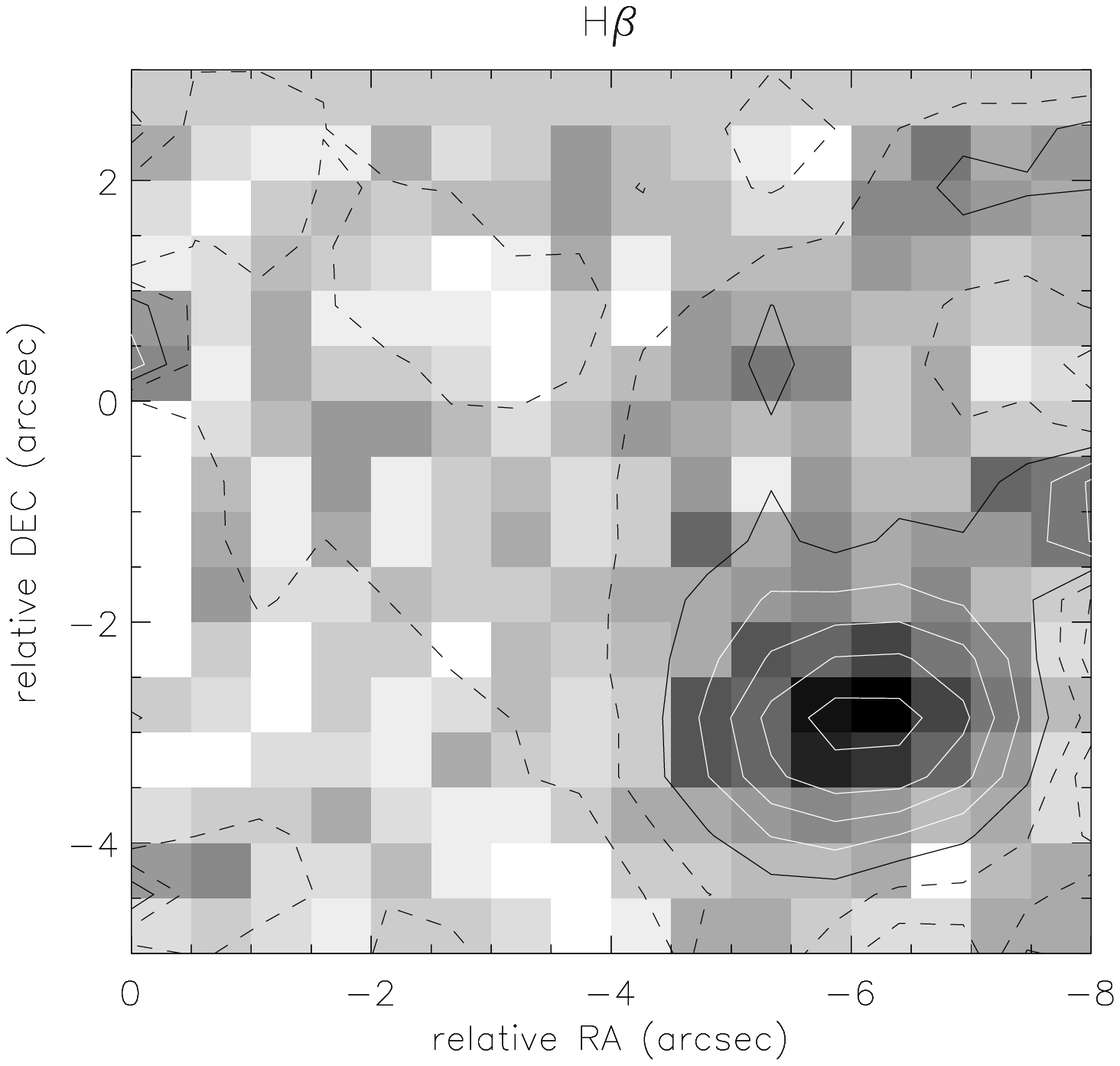}}
\end{minipage}%
\begin{minipage}[c]{.33\textwidth}
\resizebox{\hsize}{!}
{\includegraphics[]{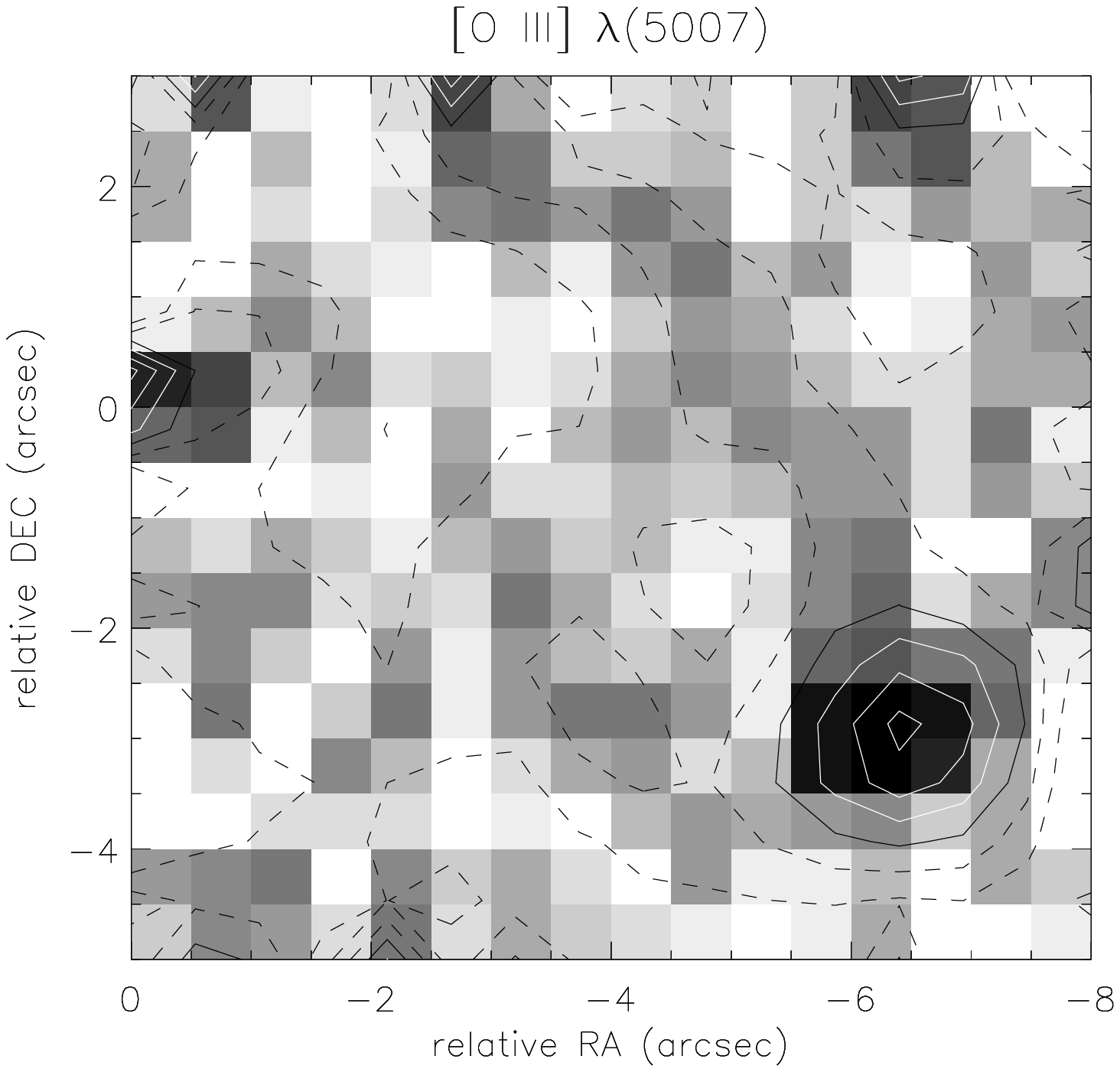}}
\end{minipage}
\\
\begin{minipage}[c]{.33\textwidth}
\resizebox{\hsize}{!}
{\includegraphics[]{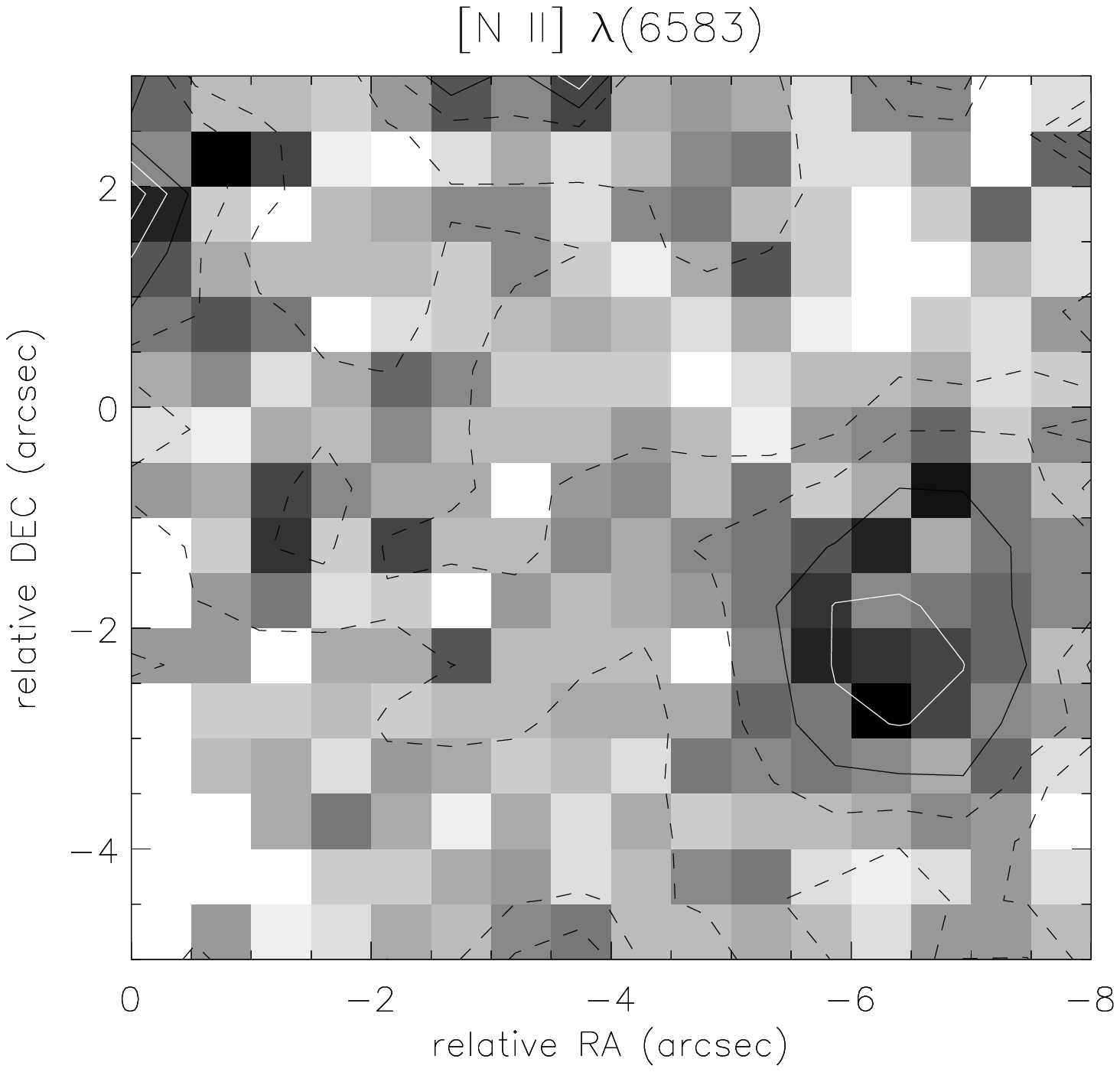}}
\end{minipage}%
\begin{minipage}[c]{.33\textwidth}
\resizebox{\hsize}{!}
{\includegraphics[]{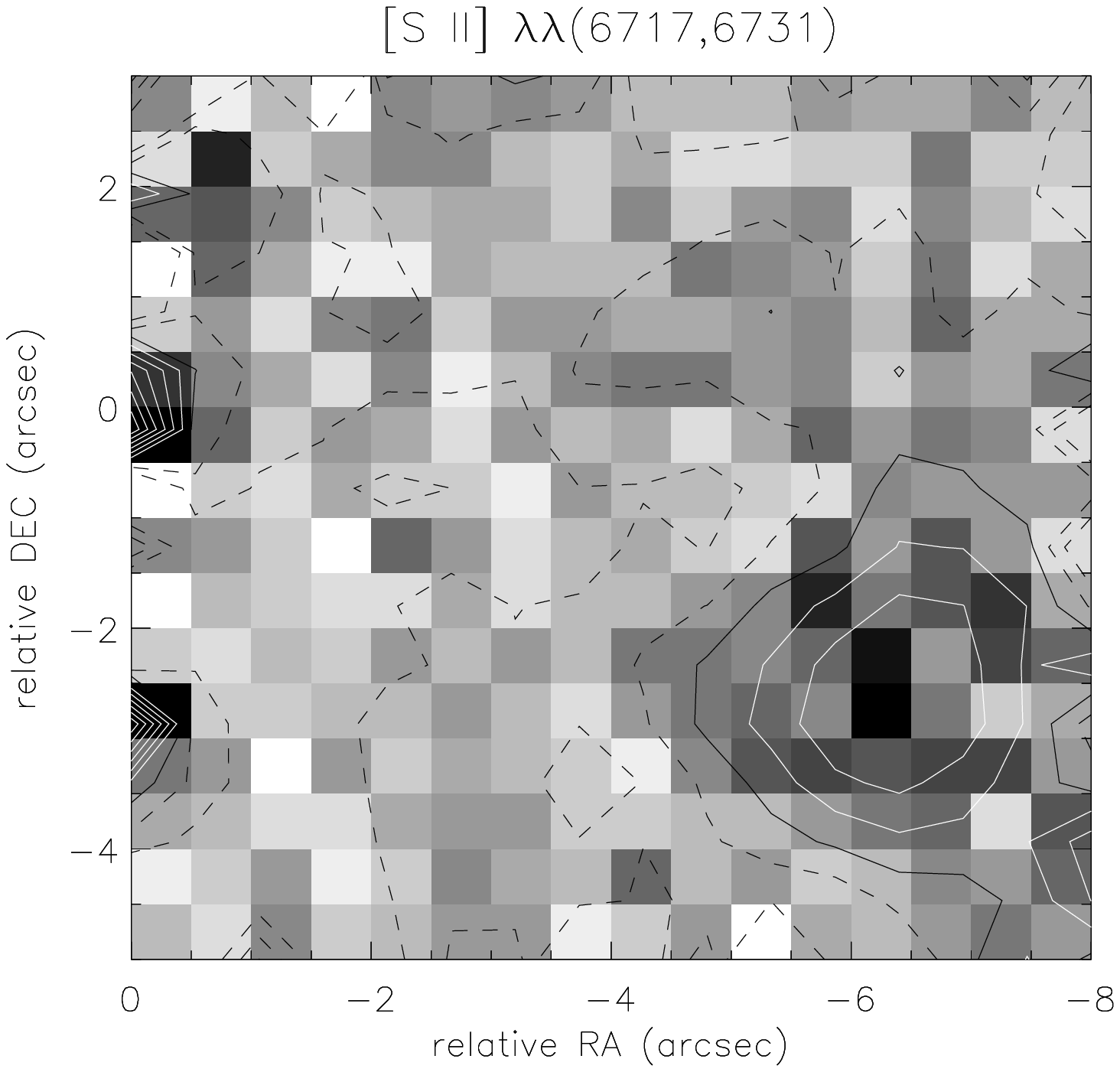}}
\end{minipage}%
\begin{minipage}[c]{.33\textwidth}
\resizebox{\hsize}{!}
{\includegraphics[]{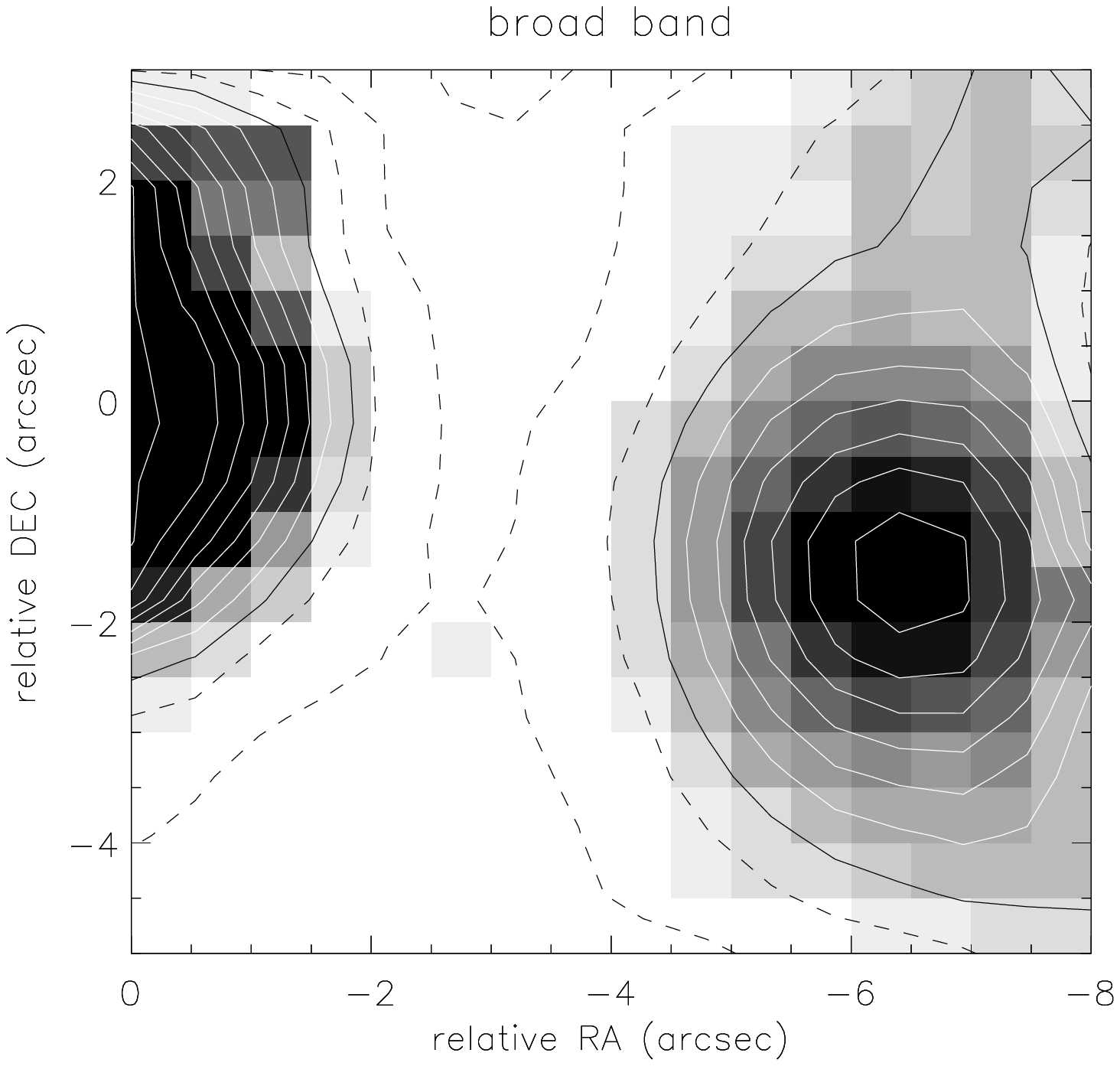}}
\end{minipage}
\caption{Emission line images of various emission lines where the continuum
  has been subtracted using images using adjacent wavelengths. For comparison
  a broad band image is included in the lower right hand panel. Smoothed
  contours of 2, 4, 6,...$\sigma$ levels are overlayed. The solid black line
  represent the 4$\sigma$ level.}
\label{fig:em_maps}
\end{figure*}

Emission line images are created by selecting appropriate narrow band filters
from the data cube. These narrow band images are created in a wavelength range
of $\pm$10{\AA} around the emission wavelength as shown in
Fig.~\ref{fig:em_maps}. The continuum emission is subtracted using narrow-band
images in wavelengths adjacent to the emission lines. Compared to the broad
band image in the lower right panel, one sees that most of the emission lines
originate $\sim$1\arcsec\ to the south of the G4 centre.  Furthermore, the
strongest emission is located in a region within a 2\arcsec$\times$2\arcsec\ 
aperture.  No other strong emission line regions are found in the field.

\subsection{High surface brightness region}
\label{high_surf}

\begin{figure*}
\centering
\resizebox{\hsize}{!}{\includegraphics[angle=90]{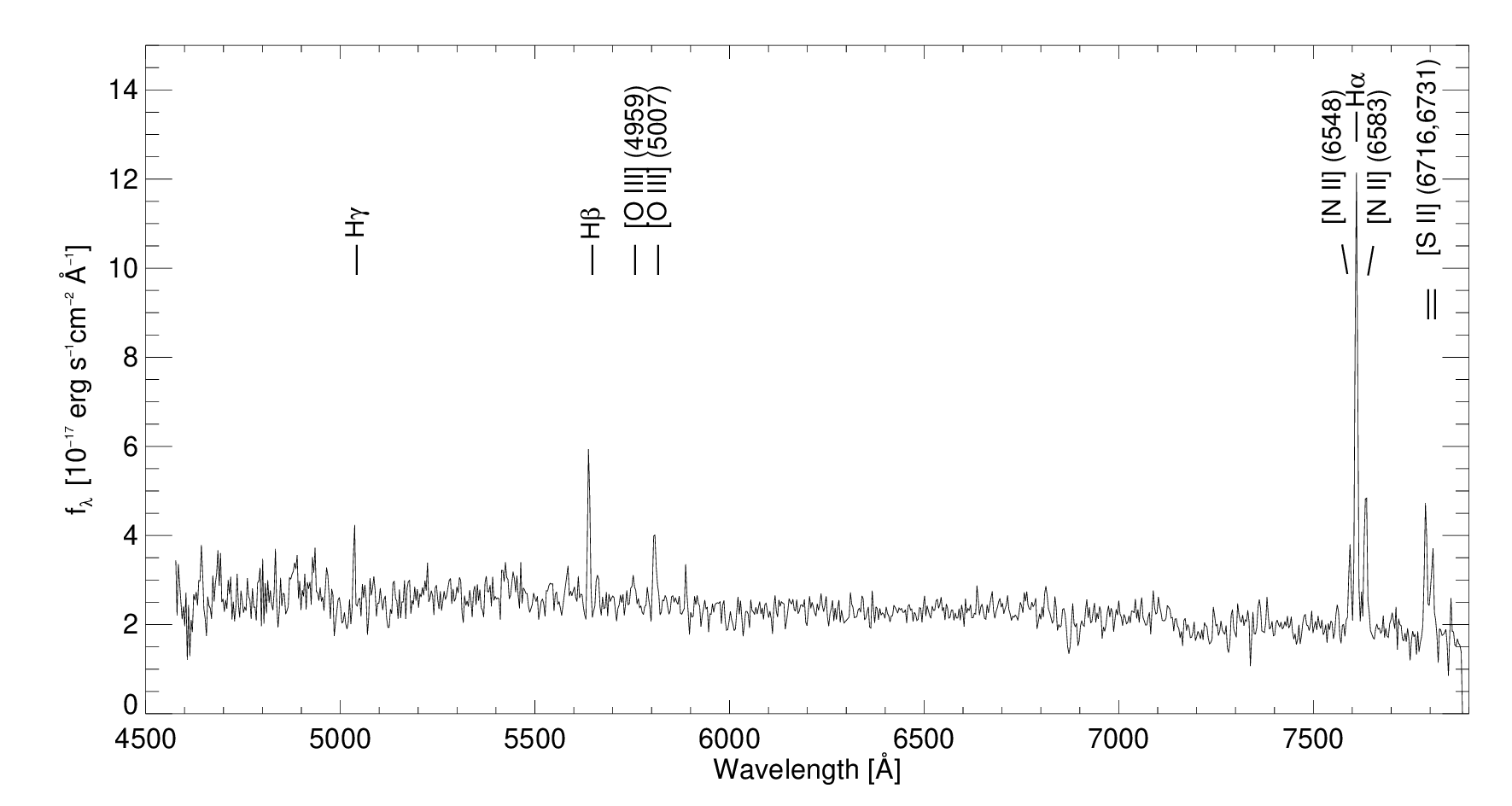}}
\caption{Spectrum of the galaxy created by co-adding 15 spaxels that
  appeared to be associated with the bright line emission
  region. Emission lines listed in Table~\ref{tab:lines} are
  indicated. }
\label{fig:gal_spec}
\end{figure*}

To create a spectrum with a higher signal to noise ratio relevant for deriving
properties from the emission lines, we selected spaxels from
Fig.~\ref{fig:em_maps} where the surface brightness of the emission lines is
strongest, i.e.  within a 1\arcsec\ radial aperture (corresponding to $\sim$15
spaxels) centered on (--2.5,--6.5). Several lines can be identified in the
spectrum shown in Fig.~\ref{fig:gal_spec}. Observed line fluxes and equivalent
widths (EW) are listed in Table~\ref{tab:lines}.   All emission line
  fluxes refer to the spectrum which has been corrected for the telluric
  absorption feature.

The median redshift of the emission lines is $z_{\mathrm{em}}=0.1595\pm0.0006$
in agreement with $z=0.1592\pm0.0020$ found by B88. The observed FWHM of
emission lines are $7.2\pm0.5${\AA} and $6.1\pm1.1${\AA} for \halpha\ and
\hbeta, respectively, which implies that the lines are consistent with being
unresolved given the spectral resolution.  Line fluxes of the faint lines have
been derived assuming a similar width as for the brighter lines. The velocity
difference between the absorption system and the median emission redshift is
180~km~s$^{-1}$.  In Sect.~\ref{kin} we will return to further analysis of the
kinematics involved in this system along with the implications.

To check if the correction for telluric absorption is appropriate we measure
the line ratio [\ion{N}{ii}]$\lambda$6583/[\ion{N}{ii}]$\lambda$6548 which
should theoretically be 3.0 \citep{wiese66}, while the derived ratio is
$3.58\pm0.76$. Thus, we conclude that the correction does not introduce large
errors of the derived line fluxes. Before application of the correction the
line ratio is $\sim$2, however, it is uncertain as the continuum level is
wrongly placed because of the telluric absorption.

Although unaffected by the correction for telluric absorption, the
[\ion{O}{iii}]$\lambda$5007/[\ion{O}{iii}]$\lambda$4959 line ratio should be
2.87, while we measure the ratio 2.31$\pm$0.85. Line ratios of the emission
lines [\ion{O}{iii}]$\lambda$5007 vs.  \hbeta\ and [\ion{N}{ii}]$\lambda$6583
vs. \halpha\ classify this object as an \ion{H}{ii} galaxy according to the
classification scheme of \citet{veilleux87}.

Using long slit spectroscopy B88 find a \hbeta\ line flux of
$2.2\times10^{-16}$ \ecs, which is slightly below our measured line flux given
in Table~\ref{tab:lines}.  Their long slit spectrum was obtained with a
1\farcs5 slit, i.e. smaller than our 1\arcsec\ radial aperture for creating
the spectrum. Therefore, our finding of a larger line flux can be explained by
small slit-losses in their spectrum.

\begin{table*}
\begin{tabular}{lllll}
\hline \hline
\smallskip
line ($\lambda_{\mathrm{air}}$)  & $\lambda^a$ & redshift & line flux & EW \\
     &   ({\AA})&   & $\times10^{-16}$ \ecs & ({\AA})\\
\hline
H$\gamma$      (4340.49)  & 5037.29 & 0.1602 & 1.53$\pm$0.48 &3.8\\
H$\beta$       (4861.36)  & 5639.60 & 0.1598 & 2.97$\pm$0.37 &8.9\\
\ion{[O}{iii}] (4958.91)  & 5754.00 & 0.1600 & 0.70$\pm$0.23 &2.8\\
\ion{[O}{iii}] (5006.84)  & 5808.78 & 0.1598 & 1.62$\pm$0.28 &6.6\\
\ion{[N}{ii}]  (6548.05)  & 7594.64 & 0.1595 & 0.83$\pm$0.17&7.6\\
H$\alpha$      (6562.85)  & 7611.88 & 0.1595 & 8.98$\pm$0.28 &43.8\\ 
\ion{[N}{ii}]  (6583.45)  & 7635.31 & 0.1595 & 2.97$\pm$0.17 &16.2\\
\ion{[S}{ii}]  (6716.44)  & 7788.96 & 0.1594 & 3.04$\pm$0.60 &17.1\\
\ion{[S}{ii}]  (6730.82)  & 7804.33 & 0.1592 & 1.95$\pm$0.50 &15.5\\
\hline
\end{tabular}
\caption[]{List of emission lines in the galaxy spectrum and the
  corresponding redshifts. Line fluxes are derived after a correction
  for Galactic extinction and telluric absorption was applied. Fluxes and EWs are the observed ones. $^a$
  Heliocentric vacuum values.}
\label{tab:lines}
\end{table*}

\section{Derived properties}
\label{prop_deriv}
Using the strong emission lines observed in the galaxy spectrum we derive the
dust reddening, abundances from the nebular emission lines in
Table~\ref{tab:lines}, and its star formation rate.

\subsection{Dust reddening}
\label{dust_red}
From the observed Balmer decrement the dust content can be estimated.  In the
Case B Balmer recombination scenario the ratio of the emission lines is
\(I(\halpha)/I(\hbeta)~=~2.85\), depending slightly on the temperature and
density \citep{brock71}, while the observed flux ratio is
$F(\halpha)/F(\hbeta)~=~3.02\pm$0.39 which is consistent with no reddening.

Using the Milky Way extinction curve in \citet{fitzpatrick99}, we find the
reddening \ebv$=0.05\pm$0.09 or equivalently $A_V=0.16\pm$0.28 for a Galactic
value of $R_V=3.1$. Similarly, the theoretical ratio of $I(\hgamma)/I(\hbeta)$
is 0.46 in the case B scenario, while we measure
$F(\hgamma)/F(\hbeta)~=~0.52~\pm~0.17$. This corresponds to \ebv=0.25$\pm$0.88
which is in agreement with the reddening derived from the \halpha/\hbeta\ line
ratio.

Because the internal extinction in this system is consistent with 0, we do not
correct the derived emission line fluxes for this effect.  With integral field
data one can in principle create dust maps by dividing the \halpha-image with
the \hbeta-image, but in this case, the signal to noise ratio is not high
enough to derive a reliable extinction map.

\subsection{Chemical abundances}
\label{abund}
Using the line fluxes of strong emission lines we derive abundances of oxygen
and nitrogen using various calibrations and diagnostics from the literature.

\subsubsection{Oxygen}
To determine the oxygen abundance of the galaxy we make use of the relation
\(O3N2\equiv \log([\ion{O}{iii}]\lambda5007 / \hbeta) / (
[\ion{N}{ii}]\lambda6583/\halpha)\) recently calibrated in \citet{pettini04}.
Our data cover all the strong lines involved in the $O3N2$ calibration which
for G4 yields $12+\log\textrm{(O/H)}=8.66~\pm~0.10$, while an additional error
of 0.14 dex is due to the scatter in the calibration itself.  The abundance
corresponds to the solar oxygen abundance 8.66$\pm$0.05 in \citet{asplund04}.
This solar oxygen abundance is lower than found in models
previously\footnote{The larger solar O abundance in \citet{holweger01}
  compared to that derived in \cite{asplund04} is due to an ignored
  contribution from a Ni blend, but also a difference between their adopted 1D
  and 3D models.}, driving the derived oxygen abundance for G4 relative to
solar towards higher values than reported in the literature for other
\ion{Mg}{ii} galaxies. Taking instead the solar oxygen abundance 8.74$\pm$0.08
obtained by \citet{holweger01}, G4 has sub-solar metallicity (0.8$Z$\subsun),
but still consistent with solar within 1$\sigma$ errors.

The $O3N2$ ratio benefits from the fact that the involved lines are not
separated by long wavelength ranges, and thus the quantity is largely
unaffected by dust obscuration. At any rate, the  small intrinsic
reddening inferred for G4 implies that the $O3N2$ ratio is affected little.

For comparison we also calculate oxygen abundances using line diagnostics
calibrated by other authors. For example, the oxygen abundance can be
estimated from the ratio $N2~\equiv\log([\ion{N}{ii}]/\halpha)$ using the
calibration in \citet{denicolo02}.  These two strong emission lines are
  present in the G4 spectrum, and the calibration gives
$12~+~\log\textrm{(O/H)}~=~8.77\pm0.06$.  Applying instead the slightly
different calibration of the $N2$ ratio in \citet{pettini04}, we find
$12~+~\log\textrm{(O/H)}~=~8.65~\pm~0.03$ in agreement with the $O3N2$
diagnostic. The intrinsic 1$\sigma$ scatter of this latter $N2$ calibration
causes an additional uncertainty of 0.18 dex.

As another line of enquiry, we combine our data of G4 with those available in
the literature. Oxygen abundances have frequently been determined using the
\(R_{23}\equiv([\ion{O}{ii}]\lambda3727+[\ion{O}{iii}]\lambda4959+[\ion{O}{iii}]\lambda5007)/\hbeta\)
intensity ratio \citep{pagel79}. Although the spectrum of G4 does not cover
the wavelength of [\ion{O}{ii}], we can estimate roughly the value if it had
been covered. If the flux reported in B88
($[\ion{O}{ii}]~=~4.1~\times~10^{-16}$\ecs) is not corrected for Galactic
extinction, the un-absorbed value will be
$f_{\mathrm{obs}}~=~4.1\times10^{0.4E_{B-V}R(5963\textrm{\AA})}$, where \ebv\ 
is the Galactic extinction, and $R(5945\textrm{{\AA}})=4.125$ is the value of
the extinction curve at the wavelength of [\ion{O}{ii}] at $z=0.1595$.
Furthermore, we correct the slit-loss present in their spectra by applying a
scale factor between their \hbeta\ line flux and ours. We estimate the
[\ion{O}{ii}] line flux
$F([\ion{O}{ii}])~=~(6.7~\pm~0.4)~\times~10^{-16}$~\ecs.  \citet{ellison04}
find an [\ion{O}{ii}] flux twice this value, but also their \hbeta\ and
[\ion{O}{iii}] fluxes are twice the ones we measure.

Using the scaled B88 [\ion{O}{ii}] line flux we find
$\log~R_{23}~=~0.48\pm$0.18, which according to the calibrations in
\citet{kobulnicky99} gives $12+\log\textrm{(O/H)}=9.02\pm0.13$. This value
agrees within the errors with the $R_{23}$-based
$12+\log\textrm{(O/H)}=8.9\pm0.2$ in \citet{ellison04} which justifies our
assumption of applying a scaling factor of the [\ion{O}{ii}] flux.
  
All the derived abundances using the various diagnostics are summarised in
Table~\ref{tab:o_abun}.  \citet{kobulnicky99} report that the calibration is
very uncertain for line fluxes with lower signal to noise ratio than 8:1.
Given such uncertainties of the latter abundance determination, combined with
the fact that the $O3N2$ calibration involves emission lines detected in the
data, and not a scaled [\ion{O}{ii}] flux derived from B88, we rely on the
oxygen abundance derived from the $O3N2$ ratio.  Thus, we find that the galaxy
has a solar oxygen abundance.
\begin{table}
\begin{tabular}{ll}
\hline \hline
\smallskip
Diagnostic     & $12+\log$(O/H) \\
\hline
$O3N2$ (1)     & $8.66\pm0.10$ \\
$N2$   (1)     & $8.65\pm0.03$ \\
$N2$   (2)     & $8.77\pm0.06$ \\
$R_{23}$ (3)   &  $9.02\pm0.13$ \\
\hline
\end{tabular}
\caption[]{Oxygen abundance determinations. Diagnostics have been taken from
  following papers: (1) \citet{pettini04}, (2) \citet{denicolo02}, (3)
  \citet{kobulnicky99}}
\label{tab:o_abun}
\end{table}

\subsubsection{Nitrogen}
Having derived a solar oxygen abundance for G4, this information can be used
to estimate the electron temperature which in turn is used to determine the N
abundance.  Because oxygen is the main coolant of a gaseous nebula, one
expects to see a correlation between the oxygen abundance and the electron
temperature of an \ion{H}{ii} region. From a sample of extragalactic
\ion{H}{ii} regions with a large range of O abundances \citet{vanzee98} derive
$T_e$.  In their 39 regions with 8.56$<$12+log(O/H)$<$8.76, similar to the
value of G4 within errors, we find an average $T_e=7000\pm900$ K.
This temperature estimate is in agreement within 1$\sigma$ errors with the
relation between $R_{23}$ and the [\ion{N}{ii}] temperature in
\citet{thurston96}, which yields $t_{[\ion{N}{ii}]}=7600\pm1000$ K. The
uncertainty both reflects the calibration scatter (500 K) and the
uncertainties of line fluxes (900 K).
  
Abundances of ionized nitrogen can then be derived using the relation between
temperature and nitrogen to oxygen ratio \citep{pagel92}.  Disregarding
ionization corrections, the nitrogen abundance can be derived assuming
(N/O)~=~(N$^+$/O$^+$), which is a valid approximation since the ionization
potentials for O$^+$ and N$^+$ are similar.  Furthermore, as shown by the
models in \citet{thurston96}, this approximation is accurate within 5\%. For
the emission lines from G4 we find $\log\mathrm{(N/O)}=-0.98\pm$0.23 using the
average $T_e$, i.e. sub-solar, but consistent within 1$\sigma$ with the solar
value $\log\mathrm{(N/O)}\subsun=-0.81$ \citep{holweger01}. A similar result
is obtained from the [\ion{N}{ii}] temperature which gives
$\log\mathrm{(N/O)}=-0.90\pm$0.22. A linear relation between the O and N
abundance at high oxygen abundances is observed in extragalactic \ion{H}{ii}
regions \citep{vanzee98b}, thus a solar N/O abundance is expected for G4.

Ionic abundances can also be derived using software for analysis of emission
line nebulae within the IRAF/STSDAS environment \citep{shaw95}. As inputs we
use the [\ion{N}{ii}] temperature estimated above and a low density
environment ($n=10$~cm$^{-3}$) which is preferred from the observed
[\ion{S}{ii}] $\lambda\lambda$6717,6731 line ratio. These values give
$\log\textrm{(N/O)}=-0.98\pm0.13$ where the error mostly depends on the
uncertainty of the temperature.  Choosing instead a density of 100~cm$^{-3}$
only increases $\log\textrm{(N/O)}$ by 0.01 dex.

As a consistency check, we investigated whether the derived oxygen abundance
is in agreement with calibrations to derive the (N/O) abundance ratio.  Using
the polynomial relations between the oxygen abundance and $\log$(N$^+$/O$^+$)
in \citet{kewley02} yields $\log\textrm{(N/O)}=-0.90\pm0.08$ inferred from the
oxygen abundance derived from the $O3N2$ diagnostics. The error includes the
uncertainty of the chosen ionization parameter. However, as shown in
\citet{kewley02} this diagnostic is relatively independent of the ionization
parameter for metallicities larger than half solar, which is the case here.
The derived abundance ratios using the various estimators are summarised in
Table~\ref{tab:n_abun}. From these values we conclude that the galaxy has
$\log\mathrm{(N/O)}=-1.0\pm$0.2.

\begin{table}
\begin{tabular}{ll}
\hline \hline
\smallskip
Diagnostic  &  $\log$(N/O) \\
\hline 
(1)  & $-0.98\pm0.23$ \\
(2)  & $-0.90\pm0.22$  \\
(3)  & $-0.98\pm0.13$ \\
(4)  & $-0.90\pm0.08$ \\
\hline
\end{tabular}
\caption[]{Nitrogen abundance determinations using different diagnostics from
  following papers: (1) \citet{vanzee98,pagel92}, (2) \citet{thurston96,pagel92}, (3)  \citet{shaw95}, (4) \citet{kewley02}. }
\label{tab:n_abun}
\end{table}

\subsection{Star formation rate}
\label{sfr}
The derived line fluxes of emission lines can be used to derive an overall
star formation rate (SFR) of the galaxy.  A redshift of $z=0.1595$ corresponds
to a luminosity distance of 2.35$\times10^{27}$~cm (760 Mpc) for the given
cosmological model. Thus, the \halpha\, line flux corresponds to a luminosity
$L(\halpha)~=~(6.25\pm0.20)\times10^{40}$~erg~s$^{-1}$. Using the relation
\begin{equation}
\mathrm{SFR} = 7.9 \times 10^{-42} L( \halpha ) \quad (\textrm{erg~s}^{-1})
\end{equation}
in \citet{kennicutt98} to convert the flux to a SFR we find
SFR~=~0.49$\pm$0.15 M\subsun \peryr. The uncertainty includes the one from the
line flux and a larger additional uncertainty from the intrinsic scatter of
the calibration of the conversion factor of $\sim$30\%.

The \halpha\ based SFR relies only on the data set of G4 presented here, while
an alternative measure of the SFR can be estimated from the scaled
[\ion{O}{ii}] line flux.  At $z=0.1595$ the luminosity is
$L([\ion{O}{ii}])=(4.67\pm0.28)\times10^{40}$~erg~s$^{-1}$, which, using the
conversion in \citet{kennicutt98}, yields SFR~=~$0.43~\pm~0.12$ M\subsun\peryr.
This value is furthermore in agreement with the calibration in
\citet{kewley04}, which yields SFR~=~$0.45\pm0.17$ M\subsun\peryr.  The
agreement between the SFR of the galaxy derived from different calibrations
gives credibility to the inferred small internal reddening in the galaxy.

\section{Kinematics}
\label{kin}
From the redshift difference between the sub-DLA cloud and the galaxy found
from the one-dimensional spectrum of G4, we find that the relative velocity
difference is 180~km~s$^{-1}$. Considering that the galaxy seems to be a
fairly luminous galaxy, such a velocity difference over a distance of 17.6 kpc
could be consistent with rotation of a massive disk, but the orientation of
the galaxy with the major axis oriented roughly north-south, while the sub-DLA
cloud lies to the north-east, does not support this hypothesis. We analyse
here only radial velocities but note that the proper motion of the cloud could
be significant.

\begin{figure}
\centering
\resizebox{\hsize}{!}{\includegraphics[]{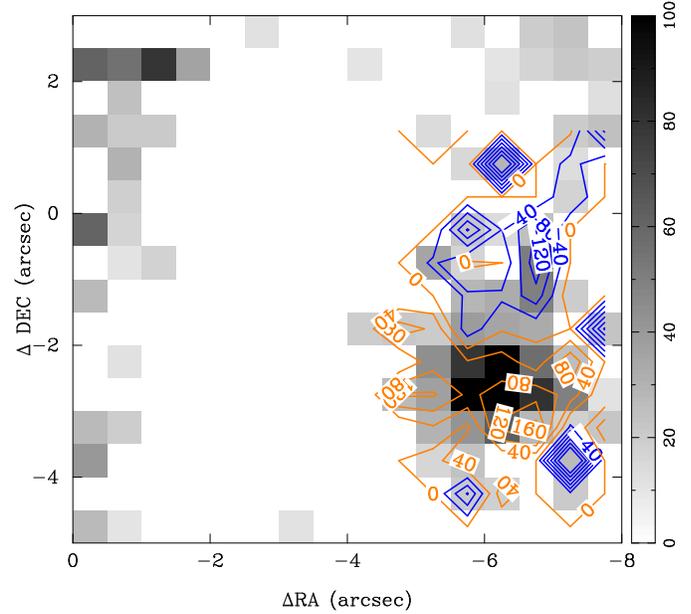}}
\caption{ The gray scale image shows the intensity of the \halpha\ emission line.  Relative velocities in km~s$^{-1}$ inferred from \halpha\ emission lines are
  shown as contours. Grey contours (orange in the {\it electronic edition})
  are positive velocities, and black (blue) contours are negative.  Only
  regions showing \halpha\ emission have been included for clarity. In
  comparison, the relative velocity difference of absorption in the sub-DLA
  cloud is +180 km s$^{-1}$ towards the QSO sight-line. }
\label{fig:kin}
\end{figure}

By fitting the strongest \halpha\ emission lines with Gaussians we examine the
dynamics of the galaxy G4. Because each individual spectrum is rather noisy,
the fitting could only be done satisfactorily in the region around the
strongest emission. In Fig.~\ref{fig:kin}, velocity offsets relative to the
systemic $z=0.1595$ are shown as contours overlayed an image of the \halpha\ 
emission intensity.  Uncertainties of the centroids of the Gaussian fits to
the strongest lines are of the order of one tenth of the spectral resolution,
i.e.  0.6~{\AA} corresponding to errors of 25~km~s$^{-1}$. We find evidence
for a systematic velocity of $\sim-120$ to $+160$~km~s$^{-1}$ with the
rotation axis oriented roughly east-west.  The point of zero-level velocity is
located roughly 0\farcs5 to the south of the centre of the G4 continuum
emission.

In this representation the absorbing cloud towards the QSO (at coordinates
0,0) has a velocity offset of +180~km~s$^{-1}$ with respect to zero velocity
of the \halpha\ emission. We conclude that it is unlikely that the absorbing
cloud is participating in a disk rotation.

In the integral field data we measure a velocity of 160~km~s$^{-1}$ relative
to the zero velocity contour of \halpha\ at a distance of $\sim$2\arcsec\ 
corresponding to 5.5 kpc.  However, if one takes the centre of the galaxy as
the true reference point, the velocity difference is $\sim$200~km~s$^{-1}$
over a distance of 2\farcs5 (6.9 kpc) as indicated in Fig.~\ref{fig:kin}.
Assuming that the galaxy is viewed edge-on, the dynamical mass is
\(M_{\mathrm{dyn}}=v_c r/G=(6\pm2)\times10^{10}\)~M\subsun, where the error is
due to the uncertainty of the velocity.  Because the inclination axis is not
known, this value is a lower limit, while another uncertainty is due to the
fact that we probably do not reach asymptotic values for the rotation curve.
In any case, the estimate shows that the galaxy is a fairly massive one.

\section{Discussion}
\label{disc}
 Having derived various properties of the sub-DLA galaxy, we proceed by
  comparing them to those of other sub-DLA and DLA galaxies in the literature.
  We emphasize once again that there is another galaxy, G3, at the same
  redshift as G4 but having a larger impact parameter. G3 might contribute to
  the absorption seen in PHL~1226. No galaxy in Fig. 3 of B88, however, is
  closer to the QSO than G4, and we discuss in section 6.3 that our data allow
  us to exclude an emission-line galaxy with a smaller impact parameter than
  G4 between G4 and the QSO. As is common practice in the literature on DLA
  galaxies, we proceed with the assumption that the absorbing cloud belongs to
  the galaxy with the smallest impact parameter.

\subsection{Relationship  with DLAs, sub-DLAs and \ion{Mg}{ii} systems}

The derived oxygen abundance of the galaxy G4 is larger than abundances
derived for high redshift sub-DLAs \citep{dessauges03,peroux03}. Furthermore,
compared to the nearby DLA absorber \object{SBS~1543+594} \citep{schulte04},
this galaxy has a higher abundance, but that goes in hand with its higher
luminosity. In fact we calculate $M_B=-20.0$ using the spectrum in
Fig.~\ref{fig:gal_all} shifted to the rest frame and convolved with the
Bessell $B$ band transmission function. G4 exhibits values which are entirely
consistent with the local luminosity-metallicity relation
\citep{kobulnicky00,pilyugin04}.

On the other hand, G4 has an oxygen abundance compatible with the upper end of
abundances derived for the \ion{Mg}{ii} selected galaxies
\citep[$-0.6<\mathrm{[O/H]}<-0.1$ derived in][]{guillemin97}. For comparison,
high resolution spectra of strong \ion{Mg}{ii} absorption systems towards a
couple of QSOs at $z\sim1$ have indicated metallicities ranging from 0.1 to 1
times solar \citep{ding03a,ding03b}. This could indicate a general agreement
between metallicities derived using absorption lines and emission line
diagnostics.

The kinematic investigation showed that the absorber is clearly not associated
with a disk, but it could be part of a rotating halo as suggested for other
\ion{Mg}{ii} absorbers \citep{steidel02}. The galaxy G4 and its sub-DLA system
could thus be similar to the $z=0.16377$ sub-DLA towards \object{Q0850+4400}
\citep{lanzetta97}. They showed that the absorption arises at large
galactocentric distance (17~kpc) and does not participate in a general disk
rotation.

Alternatively, \citet{bond01} suggested that a substantial fraction of very
strong \ion{Mg}{ii} systems arise in superwinds from galaxies. Along similar
lines, expanding superbubbles were suggested to be the explanation for metal
absorption line mirror symmetries seen in the strong \ion{Mg}{ii} absorber
towards Q1331+17 \citep{ellison03}. A superwind from the galaxy G4 is unlikely
at present time given the relatively small SFR.  However, as it is comparable
to an $L^*$ galaxy, a previous starburst could have expelled neutral gas
clouds enriched by metals.

\subsection{Abundance gradient effect}
Metallicity studies of DLAs and sub-DLAs are trying to explain the question of
what is the nature of the parent galaxy. Only in very few currently known
cases, where the parent galaxy has been clearly identified, can such an
investigation be carried out.  Abundances derived from metal absorption lines
associated with the DLAs can be compared to abundances based on emission line
diagnostics for \ion{H}{ii} regions in the galaxy. The DLA galaxy SBS~1543+594
is one example, where the impact parameter is small -- in fact, the sight line
to the QSO goes through the disk of the dwarf galaxy. In such a case, the
inferred metallicity of the DLA cloud and the galaxy is expected to be
similar.  However, in the case of the G4/sub-DLA system towards PHL~1226 where
the impact parameter is larger and the sub-DLA cloud possibly does not belong
to the disk of the galaxy, one would expect to find lower (O/H) and (N/O)
abundances of the sub-DLA cloud due to abundance gradients.

Analysing abundances in extragalactic \ion{H}{ii} regions \citet{vilacostas92}
and \citet{zaritsky94} found a large scatter in the abundance gradients for
individual galaxies of a given Hubble type.  They also found a tendency for
more shallow gradients, expressed in dex/kpc, for early and late type spirals
compared to intermediate type spirals.  Locally, three face-on disk galaxies
were found to have strong abundance gradients along the major axes in the
outer regions of their disks \citep[][hereafter F98]{ferguson98}.  Fewer
studies have analysed the metallicity gradient along the minor axes of
galaxies, but smaller metallicities have been found for extraplanar
\ion{H}{ii} regions compared to the core of an edge-on galaxy
\citep{tullmann03}. The study of \ion{H}{ii} regions in face-on galaxies can
therefore also be biased by projection effects.
 
The data presented here do not allow us to estimate the Hubble type of G4,
making it difficult to predict an appropriate abundance gradient.  If one
takes the measured metallicity gradient from F98 with an average in
$\log$(O/H) of $-0.09$~dex~kpc$^{-1}$, the abundance of the sub-DLA cloud at
17.6 kpc is expected to be $\sim$1.6~dex lower than what we find for the G4
disk.  Similarly, the average gradient in $\log$(N/O) is
$-0.05$~dex~kpc$^{-1}$ which implies an abundance ratio smaller by 0.9 dex.
Thus one could expect 12+$\log$(O/H)~=~7.1 and $\log$(N/O)~=~$-1.9$ for the
sub-DLA abundances.  We assume that the gradients are straight lines which may
not be the case \citep{zaritsky94}. If a low abundance of the PHL 1226 sub-DLA
cloud should be confirmed by future space based spectroscopy, these values
would place the sub-DLA system toward PHL 1226 among the metallicities for the
currently measured high redshift DLA and sub-DLA systems measured by several
authors \citep{pettini02,lopez02,lopez03,centurion03}.

These approximate cloud metallicities are crude estimates since the impact
parameter is a lower limit due to the unknown inclination angle, and the
individual abundance gradients in F98 vary within a factor of 2.  Furthermore,
the lowest metallicities observed by F98 reach 12+$\log$(O/H) = 7.95, thus we
are extrapolating their metallicity gradient.

\subsection{Sub-DLA cloud properties}
Other properties of G4 such as \hbeta\ luminosity, EW of \hbeta\,, absolute
magnitude, and oxygen abundance are compatible with those of emission line
field galaxies at redshifts $0.26<z<0.82$ \citep{kobulnicky03}. This
apparently seems to be in contradiction with the spectroscopic analyses
suggesting that sub-DLAs/DLAs are chemically less evolved than star forming
galaxies at similar redshifts. Specifically, sub-DLAs/DLAs have sub-solar
metallicity and element abundances suggesting low SFRs
\citep{pettini02b,prochaska03,peroux03,dessauges03}. On the other hand we can
not, with the currently available data set, exclude the possibility that a
galaxy less luminous than G4 closer to the line of sight towards PHL 1226 is
responsible for the sub-DLA.  In the integral field data presented here we
find no evidence for line emission closer to the line of sight than 5\arcsec\ 
to the west of the QSO.  With the presented single pointing towards the QSO we
can not say anything about the other directions.

As argued, either galaxy G4 or G3 could be responsible for the sub-DLA cloud.
If the cloud were associated with G3, it too has to be associated with a halo
because of the orientation of the galaxy which suggests an elongation in the
east-west direction (i.e. perpendicular to the direction toward PHL 1226).
Assuming that the sub-DLA cloud indeed belongs to the galaxy G4, we find a
velocity difference of 180~km~s$^{-1}$ from the sub-DLA cloud redshift to the
\halpha\ velocity at the centre of G4, for which we estimate the galaxy mass
$(6\pm2)\times10^{10}$~M\subsun. In this case, the escape velocity at a
distance of 17.6~kpc is 200$\pm$30~km~s$^{-1}$ implying that the cloud could
be gravitationally bound.

Yet another possibility for the location of the sub-DLA cloud is gravitational
interaction between systems. Indeed the galaxies G4 and G3 have a distance of
12\farcs6 from each other corresponding to 35 kpc, so it could be an
interacting system, but with the currently available observations we can not
test this scenario.

\section{Conclusions}
\label{conc}
Using integral field spectroscopy with PMAS we have observed the absorber G4,
previously identified by \citet{bergeron88}, toward the QSO PHL 1226 at an
impact parameter of 6\farcs4. This galaxy is most probably responsible for a
strong \ion{Mg}{ii} and sub-DLA absorption system at $z=0.1602$ in the QSO
spectrum.  At the same redshift another galaxy, G3, is present.  We cannot
determine whether the PHL~1226 absorber belongs to galaxy G4 or G3, but
concentrate on G4.  Their impact parameters of 6\farcs4 and 10\farcs9
correspond to 18 and 29 kpc, respectively, implying that either could be
responsible for the absorption given the scaling-law of \ion{Mg}{ii}
absorbers.

A strong emission line region is shown to be limited to an area of
approximately 1\arcsec\ in radius within the galaxy. In the spectra we find
emission lines from [\ion{O}{iii}], [\ion{N}{ii}], [\ion{S}{ii}] as well as
Balmer lines \hgamma, \hbeta\, and \halpha\ at the redshift 0.1595$\pm$0.0006.
We do not find regions of emission at the same redshift closer to the QSO line
of sight.
 
From the Balmer line ratios we find evidence of an intrinsic reddening of
$\ebv=0.05\pm0.16$, i.e.  consistent with 0. From the measured \halpha\ line
flux we derive a SFR=0.5~M\subsun~yr$^{-1}$.

Using the $O3N2$ line ratio diagnostics from \citet{pettini04} we derive a
solar oxygen abundance \(12+\log\textrm{(O/H)}=8.7\pm0.1\). Also a solar value
of the abundance ratio $\log$(N/O) = --1.0$\pm$0.2 is found implying a
metallicity comparable to the upper-end of metallicities for the currently
known sample of \ion{Mg}{ii} galaxies.

A kinematic analysis of the \halpha\ emission line showed that the galaxy has
rotational velocities of --120 to +160 km~s$^{-1}$ relative to the systemic
redshift with the rotational axis oriented roughly east-west. The sub-DLA
cloud, on the other hand, has a velocity difference of 180 km s$^{-1}$
relative to the galaxy and an impact parameter of 17.6 kpc above the disk
assuming that the disk is seen edge on. With such geometry and kinematics, the
sub-DLA cloud is likely part of a rotating halo and possibly gravitationally
bound. From relative velocity measurements we derive a kinematic mass of
6$\times10^{10}$~M\subsun, which corresponds to a fairly massive galaxy. The
absolute magnitude is $M_V=-20.3$ and $M_B=-20.0$ which is consistent with the
mass-luminosity relation for spiral galaxies \citep{forbes92}.

With future UV space-based spectroscopy it will be possible to compare the
metallicity of the sub-DLA cloud towards PHL 1226 with abundances derived for
the galaxy, which is necessary in order to understand the relation between the
absorption lines in QSO spectra and the galaxies responsible for them.  If
there is a difference in metallicity in line with the metallicity gradients
observed in local disk galaxies, we expect that the properties of the sub-DLA
cloud will be similar to those of high redshift DLA and sub-DLA systems. This
is an intriguing prospect which could suggest that the specific sight line
through the galaxy responsible for the DLA or sub-DLA cloud has important
consequences on the derived properties of the cloud.


\begin{acknowledgements} 
  L.~Christensen acknow\-ledges support by the German Verbundforschung
  associated with the ULTROS project, grant no. 05AE2BAA/4.
  R.~Schulte-Ladbeck is thankful for funding from HST archival grant
  no. 10282. S.~F.~S\'anchez acknowledges the support from the Euro3D
  Research Training Network, grant no. HPRN-CT2002-00305. K. Jahnke
  and L. Wisotzki acknowledge a DFG travel grant under Wi 1389/12-1.
\end{acknowledgements}
\bibliography{P_1591}
\end{document}